\documentclass[12pt]{article}

\usepackage{pdfpages}
\pdfoutput=1

\def\cl{\centerline}
\newcommand{\BEQ}{\begin{equation}}
\newcommand{\NEQ}{\end{equation}}

\def\part{\partial }

\def\nbr{\overline{n} }
\def\nb{\overline{n} }
\def\Tbr{\overline{T} }
\def\ntil{\tilde{n} }
\def\Ttil{\tilde{T} }

\def\psb{{\rho} }
\def\psbb{\bar{\psi} }

\def\np{\vfill\eject}       

\def\nb{\overline{n} }

\begin{document}

\bibliographystyle{plain}

\begin{center}
{\bf PROFILE SHAPE PARAMETERIZATION OF JET \\
ELECTRON TEMPERATURE AND DENSITY PROFILES}
\end{center}
 
\begin{center}
{\bf Beatrix Schunke} \\
{$^{ }$JET Joint Undertaking, Abingdon, Oxon, OX14 3EA, UK} \\
{\bf Kaya Imre and Kurt S. Riedel}\\
{$^{ }$ New York University, 251 Mercer St., New York  NY 10012-1185} \\
\end{center}

\begin{abstract}
The temperature and density profiles of  the Joint European Torus
are parameterised using log-additive models in the control variables.
Predictive error criteria are used to determine
which terms in the log-linear model to include.
The density and temperature profiles are normalised to their line averages
($\nbr$ and $\Tbr$).    
The normalised Ohmic density shape depends primarily on the parameter
$\nbr/B_t$, where 
$B_t$ is the  toroidal magnetic field.
Both the Low-mode (L-mode) and edge localized mode-free (ELM-free) 
high mode (H-mode)
temperature profile shapes 
depend strongly 
on the type of heating power, with ion cyclotron resonant heating
producing a more peaked profile than neutral beam injection.
Given the heating type dependence, the L-mode temperature shape 
is nearly independent of the other control variables. 
The H-mode temperature shape broadens as the effective 
charge, $Z_{eff}$, increases.
The line average L-mode temperature 
scales as $B_t^{.96}$(Power per particle)$^{.385}$.
The L-mode normalised density shape 
depends primarily on the ratio  of line average density, $\nb$,
to the edge safety factor, $q_{95}$. As $\nbr/q_{95}$ increases,
the profile shape broadens.
The current, $I_p$, is the most  important control variable for
the normalised H-mode density. 
As the current increases, the profile broadens and the
gradient at the edge sharpens. Increasing the heating power, especially the
ion cyclotron resonant heating, or decreasing the average density, peaks the 
H-mode density profile slightly.
\end{abstract}

PACS NUMBERS: 02, 52.55Fa, 52.55Pi, 52.65+z

\newpage
\section{\bf  INTRODUCTION}

Scaling law-like expressions are needed to estimate the performance of the next
generation of experimental fusion machines. While much effort has been spent on
generating scaling laws for performance parameters, 
only the Ohmic temperature profile shapes have been parameterised in terms
of the engineering variables$^{\cite{MRK,RI,IRS}}$. 
In this article, we report the results of the parameterisation of the Joint
European Torus$^{\cite{Rebut}}$
(JET) electron temperature  profiles and density shapes 
for different confinement regimes.
A preliminary version of this work was presented in Ref.~\cite{SIR}.
These parameterised
profiles summarise the typical temperature and density shapes
as  functions of engineering variables.
We use the log additive models of Refs.~{\cite{MRK,RI,IRS}}  
to represent the
profiles in terms of the normalised toroidal flux radius, $\psb$,  and the
engineering control variables such as $q_{95}, I_p ,P_{L}$,  etc.
The log-additive model is essentially a scaling expression for profile
shapes:
\BEQ \label{EQmodel}
T(\rho)/\Tbr \ \ \ {\rm or} \ \ \ n(\rho)/\nbr \ \ \ =
\mu(\rho)I_p^{f_I(\rho)} \nbr^{f_n(\rho)} q_{95}^{f_q(\rho)} \ldots \ .
\NEQ

There are at least six advantages
of profile shape scalings. First, the shape scaling summarises 
the characteristic
profile shapes over an operating period. 
Second, the fitted profiles serve as a benchmark 
against which new classes of discharges may be compared. 
Third, by fitting many discharges
simultaneously, the signal to noise ratio is enhanced and we average
over effects which are not reproducible from discharge to discharge.
Fourth, these expressions can be used in transport, stability and heating
codes as realistic temperature and density shapes. Fifth, in many
cases, physics insight can be gained from examining the profile
parameterisations. In particular, we are sometimes able to
isolate similarity variables in the profile shape dependencies.
Finally, in multi-machine databases, we can determine a size scaling
and extrapolate the profile shape to new experiments such as 
the International Tokamak Experimental Reactor$^{\cite{Post}}$ (ITER).
Thereby, our methodology can  predict the peaking factors and resulting
fusion power production. 

Our fitting methodology is described in Refs.~{\cite{RI,IRS}}.
To choose which
engineering variables to include in the fit, we minimise a model selection
criterion.
One result of this model selection procedure is that {we use different
variables to fit the Ohmic, low mode (L-mode) and high mode (H-mode) cases.}
We apply a similar sequential variable selection procedure in our
linear regression analysis of the line average temperature scaling.
To the best of our knowledge, sequential variable selection has never 
been used in global confinement analysis, and as a result, confinement
scalings often include unnecessary and insignificant control variables.  
By using only the important control variables, we imply that the other control variables
barely influence the profile shape in our database. Thus, {\em we find the 
parameter directions where the profile shape is invariant.}

\newpage

Our list of candidates for control variables  includes
engineering control variables    
such as
the logarithms of the
edge safety factor, $q_{95}$, the plasma current, $I_p$ (in MA), the
toroidal magnetic field, $B_t$ (in Tesla),
the average density, $\bar{n}_e$ (in $10^{19}/m^3$), and the
 loss power, $P_L \equiv P_{aux}+P_{Ohm} - \dot{W}$ (in MW).
We find that ion cyclotron resonant heating (ICRH) yields  profile
shapes different from the neutral beam injection (NBI) heating shapes. 
To measure this 
difference, we define a new  variable, the heating fraction, $H_{fr}$, 
to be the ratio of ICRH to total auxiliary heating: 
$H_{fr} \equiv P_{ICRH}/[P_{NBI} +P_{ICRH} ]$. 
We also examine for possible dependencies on
the effective ion charge, $Z_{eff}$,
and the plasma
inductance, $\ell_i$. 
We normalise the variables about their mean values in the data set.
We do not consider the major and minor radii ($R$ and  $a$)  
and the plasma elongation, $\kappa$,
because of their small variation
in the data set. To determine the size dependence and examine intra-machine
variability, 
a multi-machine database with an expanded  parameter space is necessary.
Work on a combined DIII-D-JET profile parameterisation 
has begun$^{\cite{SSIR,IRSS}}$. 

To measure the goodness of fit and determine which control variables 
influence the profile shape, we use the
Predictive Absolute Residual (PAR) criterion as defined in the appendix.
{\em The PAR criterion is our estimate of the expected absolute error
in predicting the normalised profiles of new data taken under similar
operating conditions.}
The PAR criterion is the mean absolute residual error
with a degree of freedom correction. 
The residual fit errors tend to be more uniform on an
absolute scale than a logarithmic scale and appear to be nearly independent
of the reported error bars.  
Thus we replace the Rice criterion 
(which assumes the errors are proportional to the reported error bars
on a logarithmic scale) of Refs.~{\cite{RI,IRS}}
with the PAR criterion which assumes the errors
are uniform on an absolute scale.
Using the sum of absolute residual errors instead of the sum of 
squared errors robustifies the criterion to outliers.  

The choice of model selection criteria is somewhat arbitrary
for real tokamak data because the data departs in unknown ways from the 
idealistic assumptions on which these criteria were based. We believe
that the PAR criterion  is robust and reasonably represents the size
of the residual errors.
Nevertheless, the exact choice of when to truncate the model and quit
adding new control variables to the model is an art. As a rule, 
we quit adding terms
when the PAR value only decreases slightly with the addition of
a new term, and when several different control variables yield
nearly the same PAR value.
The PAR criterion and the Rice criterion usually
agree on which control variables are important to keep in the fit.
The Rice criterion sometimes wants to keep one more control variable 
in the model.

In addition to replacing the Rice model selection criterion with
the PAR model selection criterion, we make three other modifications
of the fitting methodology in Refs.~{\cite{RI,IRS}}:

First, we normalise each temperature profile to its line average.
(This is not an issue for the density profile since $\nbr$ is a control
variable.) We estimate the profile shape and the line average temperature
separately.
When we did not normalise the profiles, the errors in predicting
the line average dominate the fit error. Normalising the profiles
removes the uncertainty in the line average temperature, $\Tbr$, 
and allows a more accurate parameterisation of the profile shape. 
To show that we are not losing information by fitting the normalised profiles 
and the line average temperature separately, we allow the shape model to 
depend on  $\Tbr$. We find that using $\Tbr$ as a predictor variable
does not improve the profile shape parameterisation.

Second, we use only the measured data on the outboard flux radii and
symmeterise the fit about $\rho=0$. This change is based on the
assessment of the relative reliability of the inboard and outboard 
measurement locations by the JET diagnostic team. 

Third, we increase the spline penalty functional and thereby the smoothness 
of the fitted curves.  Traditional data adaptive smoothness  criteria 
attempt to minimize the predictive fit error in a particular function
space. Surprisingly, minimising the  predictive fit error results in
curve estimates that tend to have false inflection points and spurious
wiggles. These spurious wiggles can be eliminated  with high probability
if larger smoothing is used. We believe that achieving a correct estimate
of the shape (correct number of inflection points) is more important
than minimizing the model selection criterion with respect to the smoothness
parameters. Thus we inflate the smoothness parameter by an amount which
guarantees that asymptotically the shape of the estimate will be correct.
This modest increase in the smoothness parameter has little impact on 
the final value of the residual fit error.
For a more theoretical analysis of shape correct fits, we recommend 
Ref.~{\cite{RiedelPCC}}.

Our predicted profiles are not ``dimensionally correct'' in the sense
that they do not impose the Maxwell-Boltzmann constraint. Our predictions
may be made dimensionless by adding a suitable exponent of the form
$(R/R_{JET})^{f_R(\rho)}$, where  ${f_R(\rho)}$ is determined by the
linear constraint. (See Ref.~\cite{RiedLM} for the constraint procedure.)
Our research$^{\cite{RiedHM}}$ indicates that 
the H-mode confinement violates  
the Maxwell-Boltzmann constraint, probably indicating the key role of radiation
loss in many H-mode discharges.

In the fitting procedure, we do not explicitly require that the
predicted profiles have a line integral equal to one. 
(We only normalise the data
to correspond to profiles with  a line integral equal to one.)
Implementing the profile normalisation in the fitting procedure would
require much programming effort and computational cost. Not normalising
the fitted profiles introduces a 2\% error in the predicted line
averages for the database. This error can be much more when our models
are used to predict new profiles for parameters outside of the database range.
To minimise these nonlinear errors, we recommend 
that our profile predictions be normalised to have a line integral to one.

\
 

\noindent
\section{\bf  DATA DESCRIPTION}



We use databases of between 44 to 52 discharges.
The discharges were taken from the experimental campaigns in 1989-92.
The Ohmic and  L-mode
discharges are mostly limiter discharges with a beryllium-evaporated wall.
 Tables \ref{LDB} and \ref{HDB} summarise the L-mode and H-mode databases.
The elongation, $\kappa$, varies by only 7.3\% for H-mode
and by  only 9.9 \% for L-mode.
The typical value of the auxiliary heating differs considerably
in the two cases. For the  (H-mode database, the mean is 8.0 MW while it is
only 4.8 MW for the L-mode database. 
This difference is due to the H-mode power threshold.

The H-mode discharges are typical H-mode discharges from
1989-92, and therefore are predominately edge localized mode-free (ELM-free). 
Our method assumes that the plasma profiles are time stationary.
A significant number of H-mode profiles are still evolving to 
a limited extent, and thus our H-mode fit results may be
influenced by the ongoing  temporal evolution of the profiles.

The flux radius is  normalised such that the toroidal flux through
a given radius, $\psb$, is equal to $\psb^2$ times the total flux.
In our previous articles$^{\cite{RI,IRS}}$, 
we used the normalised poloidal flux radius,
$\psbb$, instead of the normalised toroidal flux radius.
For the line average, we use $\nb \equiv 
 \int_{0}^{1} n(\psb)d\psb$ instead of the more common definition
of $\nb \equiv  \int n(R)dR / \int dR$.
For each profile, we calculate the line average electron density
from the  LIght Detection And Ranging
(LIDAR) diagnostic measurements 
instead of using the interferometry measurement.
We use the line averages of the temperature and density instead of the volume
averages, however this choice is done for convenience only. We expect
similar results to hold for the volume averages.

The electron temperature and density profiles are measured by the JET LIDAR
Thomson scattering diagnostic$^{\cite{LIDAR}}$.
Each profile is measured at approximately
50 radial locations along the plasma mid-plane.
As discussed in Ref.~{\cite{IRS}, neither the profile measurements 
nor the accuracy of the measurements are  symmetric 
with respect to $\rho$. 
The outboard side measurements are more accurate than the inboard side.
Thus, we use only the outboard side measurements, 
and we require that the fitted functions be symmetric in $\rho$.
We find that 10 internal knots are adequate to describe the shape. 
The LIDAR measurements usually have higher edge temperatures than
the ECE measurements on JET. Our temperature profile parameterisation 
represents the LIDAR measurements.
 
Our present results supercede our earlier results in 
Refs.~\cite{RI,IRS}.


\noindent
\section{\bf OHMIC PROFILE FIT}

\subsection{\bf Ohmic Density}

The Ohmic database 
is described  in Ref.~3. Table~\ref{ODB} summarises the variation
of the control variables.
We 
normalise the density by its line
average,  $\nb$: $n(\psb)/\nb$. 
Normalisation of the profiles greatly reduces the fit error.
We define $y_{i,j} \equiv \ln[n_{i,j}/\nb_i]$, where $n_{i,j}$ is the measured
electron density of the $i$th profile at the $j$th radial location.

We begin by considering all possible two-term fits to the data of the
form:  $y_{i,j} = f_0(\psb_j) \ +\ f_1(\psb_j) \ln[u_i]$, where
$f_0(\psb_j)$ and $f_1(\psb_j)$ are spline functions and $u_i$ is
the value of the control variable, $u$ for the $i$th profile.
The first column of Table \ref{ODPAR} is a list of control variables,
where the logarithmic transformation is implicitly assumed.
The second 
column is the PAR goodness of fit when a single control variable
is used in the fit. Of the standard control variables 
(not including $\nbr/B_t$), the most effective in reducing the PAR statistic
is the average density, $\nbr$. 
The third column 
gives the PAR goodness of fit criterion
when we use two control variables in the fit,
the first of which is  $\bar{n}$. 
Using both $\nbr$ and $B_t$ reduces the PAR statistic from
.0612 to .0567.
 
The fourth 
column  shows that the goodness of fit
is not improved by fitting the data with three control variables
($\nbr$, $B_t$, and one other variable).
Figure \ref{ODmu} plots $\mu ( \psb) = \exp (f_0 ( \psb) )$, $
f_{n} ( \psb)$
and $f_B ( \psb )$. Here $\mu ( \psb)$ is our predicted profile
when $\nbr$ and $B_t$ are chosen at their geometric means.
Thus  $\mu ( \psb)$
corresponds to the canonical normalised Ohmic density profile. 
At $\rho = .75$, the gradient of $\mu(\rho)$ becomes steeper.
Since $f_n(\rho)$ is negative in the interior and positive 
beyond $\rho = .4$, our predicted profile broadens as 
the average density increases.

The most striking result of Figure \ref{ODmu} is that 
$f_n(\rho) \approx - f_B(\rho)$. To very good approximation,
the shape depends on $\nbr$ and $B_t$ only through 
the ratio, $\nbr/B_t$. Since the major radius is fixed in JET,
 $\bar{n} /B_t$ is a Murakami-like variable.
The Murakami parameter is  associated 
with increased radiation near the density limit.
This offers some potential insight into the physics of the
Ohmic density profile variation being related to radiated energy loss.

The slight difference
between the two curves ($f_n(\rho)$ and $- f_B(\rho)$)
near $\rho = 1.0$ indicates that increasing 
both quantities equally will lead to a slight broadening of the edge.
We do not believe that this difference is significant.
To show this, we refit the model using only the ratio,  $\nbr/B_t$,
as a control variable. The resulting PAR value, .0561, is lower
than using both $\nbr$ and $B_t$ separately. This shows that
using two free functions is not worth the cost of the extra degrees
of freedom. 

The final  column of Table \ref{ODPAR} examines the  fit 
when two control variables
are used, one of which is $\nbr /B_t$.
The goodness of fit is  not  improved significantly
by adding a third control variable.
Therefore, we adopt the one control variable model
using only $\nbr/B_t$:
\begin{equation}\label{E2a} 
n ( \psb ) = \bar{n} \  \mu( \psb )\
\left({\bar{n} \over B_t} \right)^{ f_{n/B} ( \psb )}
\ , \end{equation}
where we assume that $\bar{n} \over B_t$ has been normalised to its
mean value.
Figure~\ref{ODdat} plots our predictions versus the data for the profile  with
the largest value (1.28) of $\nbr/B_t$ and for the profile
with the smallest value (0.44).  At  $\nbr/B_t= 1.28$, our predicted profile
is just beginning to be hollow. The fitted profiles do not attempt to 
track the slight flattenings in the data that are due to either 
random measurement noise or possibly nonreproducible magnetic islands. 
Instead, the fitted profiles model the diffusive part of the profiles
and average over the local flat spots.


\subsection{\bf Ohmic Temperature}

Our previous Ohmic temperature fit is described in detail in Ref.~3. 
To a reasonable degree of accuracy, the JET Ohmic 
electron temperature profile can be fit with the ``profile resilient'' form:
\begin{equation}\label{OTE1}
T(\psb) = \mu (\psb)  q_{95}^{f_q (\psb)}\ 
\end{equation}                 
A more accurate fit to the data is adding a magnetic field dependence: 
\begin{equation}\label{OTE2}
T(\psb) = \mu (\psb)   q_{95}^{f_q (\psb)}\ B_t^{f_B (\psb)} \ ,
\end{equation}                 
which reduces the PAR value from .0812 to .0755. 
The final column of Table \ref{OTPAR} shows that little 
improvement in the fit occurs
when a third variable is added. We believe that the dependence
on plasma elongation is not real due to the small amount of $\kappa$ variation
in the data set. Instead, we believe that the observed
 $\kappa$ dependence
is due to hidden variables (differences in machine operation) 
which correlate with
$\kappa$ variation in this data set. Thus we accept the two variable model
of Eq.~(\ref{OTE2}).

The final row of Table \ref{OTPAR} uses the line average temperature
as a regression variable. Since the fit barely improves when $\Tbr$
is used, this demonstrates that the temperature shape and magnitude
are uncorrelated.

Figure \ref{OTmu} plots $\mu ( \psb) = \exp (f_0 ( \psb) )$, $
f_{q} ( \psb)$
and $f_B ( \psb )$. Here $\mu ( \psb)$ is our predicted profile
when $q_{95}$ and $B_t$ are chosen at their geometric means.
The canonical normalised Ohmic temperature profile,  $\mu ( \psb)$,
is bell-shaped. Increasing $I_p$ and $q_{95}$ corresponds to 
changing the shape according to $f_q(\rho)$. Higher $q_{95}$ and
$I_p$ result in a more peaked profile. From our previous study$^{\cite{IRS}}$,
we believe that changing $q_{95}$ with fixed $I_p$ and fixed $B_t$ should
change the profiles less. 
This $\hat{q} \equiv q_{95}I_p/B_t$ effect is not in our present model.

When $q_{95}$ is increased by increasing both $B_t$, the profile
shape change corresponds for $f_q(\rho)+ f_B(\rho)$. Thus changes in $B_t$
cause little effect near the center, but very sharp changes in the edge 
gradient. 
Figure~\ref{OTdat} plots our predictions versus the data for the profile  with
the largest value (12.6) of $q_{95}$ and for the profile
with the smallest $q_{95}$ (2.88)  in the data set.  
The local flat spots are not described by our model when they are not
reproducible. Instead, the fitted profiles average over the local flattening 
and correspond to the net diffusive profile.

In Eq.~(\ref{OTE2}), the $\Tbr$ dependence is unknown. We fit the 
line average temperature with a power law in the control variables
using a log-linear regression. 
We find that a three-parameter model using $I_P$, $\nbr$ and $B_t$
fits the line average temperature:
\begin{equation}\label{OTbr}
\Tbr    = .505 \ I_p^{.64\pm.04}\ B_t^{.54\pm .09}\ (\bar{n} )^{-.31\pm.06}
\ . \end{equation}
The degree of freedom corrected root mean square error (RMSE)
is 11.5 \%.
The most surprising part of the $\Tbr$ regression is that
$\Tbr \sim  (\bar{n} )^{-.31}$ instead of the often assumed scaling of
$\Tbr \sim  (\bar{n} )^{-1.0}$. In Ohmic plasmas, the heating rate
is coupled to the line average density so it is not surprising
that $\Tbr \nbr \ne\ constant$. Our L-mode and H-mode results
confirm that the line average temperature has a relatively
weak density dependence in JET. In Ref.~\cite{MRK}, Ohmic 
results for the axisymmetric  divertor experiment$^{\cite{ASD}}$ (ASDEX) 
are given: 
$<T_e^{ASDEX}> \sim  I_p^{.95}\  B_t^{.04}\ (\bar{n} )^{-.56}$,
where we have averaged the two ASDEX scalings.
Thus JET has a weaker density dependence, a different $q_{95}$
dependence and a stronger magnetic field dependence. All of these
differences in the $\Tbr$ scaling are consistent with the
differences in confinement scalings$^{\cite{RiedLM}}$.
One of the important results of Ref.~\cite{MRK} is that the plasma
energy shows no roll-over regime at high density. The  flattening
in confinement time in ASDEX correlates with an increase in
the loop voltage, allowing the plasma energy, $W$, to 
scale similarly in both regimes. Similarly, we suspect that the
$\Tbr$ will change only slightly in the roll-over regime.

\noindent
\section{\bf  L-MODE PROFILES}
\vspace{.15in}

\subsection{L-mode Density}
\vspace{.15in}

In analyzing auxiliary heated discharges, we add two more control
variables: the  loss power,
 $P_L \equiv P_{aux}+P_{Ohm} - \dot{W}$ (in MW), and the heating fraction,
$H_{fr} \equiv P_{ICRH}/[P_{NBI} +P_{ICRH} ]$. 
We find that ion cyclotron resonant heating (ICRH) produces different profile
shapes than neutral beam injection (NBI) heating produces. We believe that
this difference is due to changes in the power deposition profile
and changes in the fast particle population. In the future, we hope
to examine more physical parameterizations of the heating 
effectiveness$^{\cite{CalCord}}$.
 For this study, we use the heating fraction because
it is easily evaluated. 
In the L-mode data set, 28 of the discharges are predominately 
(more than 50 \%)
ICRH-heated and 24 are predominately NBI-heated.

Table \ref{LDPAR}  
presents the goodness of fit as the number and choice of control variables
is varied.
The most important control variable is the line average density, $\bar{n}$.
The first column of  Table \ref{LDPAR} shows that using only $\nbr$ yields
a PAR of value of .0633. If $\nbr$ is replaced by $\nbr/B_t$, the misfit 
increases to a PAR of .0641. 
As seen from Fig.~\ref{LDmu}, $f_n(\rho)$ is shaped like a convex parabola. 
This implies that
as the average density increases, the profile broadens. 
It is reassuring that our statistical model
agrees with this common experimental observation.

The second most important variable (given that $\nbr$ is used)
is the  edge safety factor, $q_{95}$. The two control variable model
has a PAR value of $0.0619$. The $f_q(\rho)$ curve in Fig.~\ref{LDmu} is 
shaped like a concave parabola. Note that $f_q(\rho)  \approx - f_n(\rho)$.
Thus increasing both the edge $q$ and the density by the same relative
amount should result in little change in the profile shape.
The approximate relation,  $f_q(\rho)  \approx - f_n(\rho)$,  
tempts us  to replace the two-parameter fit
with a new model involving only the ratio, $\nbr/q_{95}$.

The third column of Table \ref{LDPAR} shows the goodness of fit 
for three control variable models given $\nbr$ and $q_{95}$.  
The addition of the heating fraction,
$H_{fr}$ reduces the PAR value to .0606. From Fig.~\ref{LDmu}, 
increasing the fraction
of ion cyclotron heating tends to broaden the density profile.
This may indicate that ICRH tends to increase the inward pinch of particles.
Alternative explanations are discussed in the L-mode temperature section.  

Column 4 of Table \ref{LDPAR}  shows that the adding additional variables 
does not appreciably improve the residual error. Thus, we select
the L-mode model:
\begin{equation}\label{E3a}
n(\rho) =\bar{n} \  \mu(\rho) \
( \bar{n})^{f_{n} (\rho)} \
q_{95}^{ f_{q}(\rho)}\   
\exp\left(f_H(\rho)H_{fr} 
\right) \ .
\end{equation}
Figure~\ref{LDdat} plots our predictions versus the data for the profile  with
the largest value ($5.63\times 10^{19}/m^3$) of $\nbr$ and for the profile
with the smallest value ($0.78 \times 10^{19}/m^3$).  
Since we fit 52 profiles simultaneously, the predicted curves follow
the reproducible part of the profile and neglect the local flattenings
which vary from discharge to discharge.


\subsection{L-mode temperature}     

The model selection criterion (PAR) shows that the heating fraction,
$H_{fr} \equiv P_{ICRH}/[P_{NBI} +P_{ICRH} ]$ 
is the only control variable that
is necessary to parameterise the L-mode temperature profile shape.
The first column of Table \ref{LTPAR} shows that the PAR value is
much smaller when $H_{fr}$ is used than with any other single control variable.
The second column shows that the goodness of fit does not improve if a second 
control variable is added to the model.  Using three control variables,
$H_{fr}$, $P_L$ and $I_p$, reduces the PAR value insignificantly from
.0869 to .0868. Our judgement is that this reduction in fit 
error is insufficient to justify adding two additional control variables
to the model. Therefore, we recommend the one-parameter model:
\begin{equation}\label{EqLT}
T(\rho) =\bar{T} \  \mu(\rho) \ \exp\left(f_H(\rho) H_{fr}
\right)
\  .
\end{equation}
Figure \ref{LTmu} shows $\mu(\rho)$ and $f_H(\rho)$. 
Here, $\mu(\rho)$ corresponds to
the database mean of the normalised temperature profile. Thus, the
canonical profile $\mu(\rho)$ has a bell shape. 
The heating factor function, $f_H(\rho)$, shows that increasing the percentage
of ICRH power to NBI power results in increasingly peaked profile shapes.
This shape dependence could be due to the different power deposition profiles
of ICRH and NBI. In Ref.~\cite{CalCord}, 
it is shown that the temperature profile
shape varies as the resonance layer for ICRH is shifted outward.
Since the NBI power deposition profile is broader than the typical
ICRH deposition profile, our shape parameterisation is consistent with
the results in Ref.~\cite{CalCord}. 
Another factor in the heating type dependence is 
the reduction of sawtooth activity with ICRH. The underlying physical
mechanism is that fast particles stabilise the sawtooth at least partially,
thereby reducing  the sawtooth frequency.

In our previous analysis 
(using the poloidal flux, $\psbb$, instead of the toroidal flux, $\rho$),
the toroidal magnetic field influenced the profile shape. 
Using the normalised toroidal flux radius as the spatial coordinate
makes the profile shape nearly independent of the magnetic field
magnitude. Note that the mapping $\psbb$ to $\rho$ depends on the
safety factor, $q(\rho)$.

In Eq.~(\ref{EqLT}), the $\Tbr$ dependence is unknown. We fit the 
line average temperature with a power law in the control variables
using a log-linear regression. Since not all control variables are
necessary to fit $\Tbr$, we again use a sequential selection procedure
to choose which variables are important to include in the regression.
We find that a three parameter model using $P_L$, $\nbr$ and $B_t$
fits the line average temperature:
\begin{equation}\label{E4}
\Tbr    = .705 \ B_t^{.97\pm.16} \ (\nbr)^{-.38\pm.07} \  P_L^{.36\pm .05}
\ .
\end{equation}
The corrected RMSE is 20.0 \%, indicating that the power law model
may not be appropriate for the line average temperature.
Equation (\ref{E4}) can be interpreted as {\em Temperature $/B_t \sim$
(Power per particle)$^{.4}$, which is a gyro-Bohm scaling.}
Thus we see that the electron temperature is gyro-Bohm while the
global confinement time is Bohm-like.
The main surprise of our $\Tbr$ fit is that there is no dependence
on the plasma current in our fitted expression. Note that $I_p$
varies by a factor of four; thus the lack of a current dependence is highly
significant. Most tokamaks tend to observe a  dependence
such as $I_p^{.85}B_t^{.2}$ for the L-mode confinement time.
For JET, the L-mode confinement$^{\cite{RiedLM}}$ tends to scale as 
$\tau_E \sim I_p^{.9} B_t^{.5} $, which shows more of
a $B_t$ dependence. 
Equation (\ref{E4}) indicates that the improvement in L-mode
confinement time with increasing $B_{pol}/B_t$ is not due to 
improved confinement of bulk electrons. Thus the $I_p$ dependence 
of the $\tau_E$ scaling is associated with improved confinement 
of ions or fast particles or possibly due to a reduction in $Z_{eff}$
and the corresponding increase in ion density.

Figure~\ref{LTdat} plots our predictions versus the data for a profile  with
only NBI heating and for a profile with only ICRH. For this particular
ICRH-only profile, our prediction is less peaked than the measured profile.
Nevertheless, the prediction is easily within $2\sigma$.


\noindent
\section{\bf  H-MODE PROFILES}

Our H-mode data set consists predominately of ELM-free
discharges, that were typical of H-mode operation in the period 
1989-92. Many of these discharges are evolving to
a greater or lesser extent. This nonstationary behavior
makes our H-mode profile parameterization somewhat
less reliable than our corresponding Ohmic and L-mode
parameterisations.  
In the L-mode data set, 28 of the discharges are predominately 
(more than 50 \%)
ICRH-heated and 20 are predominately NBI-heated.

\subsection{ H-mode Density}


Figure \ref{HDmu} displays the predicted profile at
geometric  mean of the normalised density profiles.
The canonical ELM-free H-mode profile is much flatter than the corresponding
canonical L-mode profile (Fig.~\ref{LDmu}). 
Even at $\rho=1.0$, the local density
is more than 60 \%  of the central density.

The total current is the most important control variable in reducing
the model selection criteria. Increasing the total current, $I_p$,
flattens the normalised density profile. As a result, the edge 
gradient steepens.
At the largest values of $I_p$, the density profile is often hollow.
The second most important control
variable is the  loss power, $P_L$.
Figure \ref{HDmu} shows that increasing heating power  broadens the profile
in the central region. In contrast to the current dependence,
increasing the input power does not lead to  
large changes in the edge gradient.
This effect in seen in Figure \ref{HDmu}, since $f_P(\rho)$ is smaller than
than $f_I(\rho)$ near $\rho=1$. 
Table \ref{HDPAR}   presents the model selection criteria. The PAR statistic
decreases from .0619 to .0597 when $P_L$ is added to the model.
From the third column of Table \ref{HDPAR}, we see that adding either
the line average density or the heating fraction reduces the PAR
statistic to .0584. The fourth column shows that the PAR value
reduces to .0563 when both  
the line average density and the heating fraction are added.

Figure \ref{HDmu} shows that increasing the average density broadens the 
density profile but does not effect the edge gradient.
Increasing the heating fraction of ICRH results in somewhat more peaked H-mode 
density profiles. The opposite heating fraction dependence is observed
in L-mode, where ICRH broadens the density profile.
Our final H-mode density model is
\begin{equation}\label{E62a}
n_e(\psb) = \nb \ \mu (\psb ) \ I_p^{ f_I (\psb )}
 \ P_L^{f_{P} (\psb)}\ (\nb)^{f_{n} (\psb)} \exp\left(f_H(\rho)H_{fr}
\right) \ . \end{equation}
Figure~\ref{HDdat} plots our predictions versus the data for both the profile
  with the largest value (3.18) of $I_p$ and  the profile
with the smallest value (2.07) in the data set.  
The fit quality is good for the $I_p = 3.18$ case, while the $I_p = 2.07$
case is more peaked than our prediction. This small misfit occurs
because other low current discharges in our database are much less 
peaked than \# 27215. The predicted profiles show a local flattening
near $\rho = 0.9$. This feature is present in a large percentage of the
data set and therefore, our model tracks this reproducible flattening.




\subsection{H-mode Temperature}

The database mean of the normalised temperature is seen in Fig.~\ref{HTmu}.
The ELM-free H-mode  canonical profile shape is broader 
than the corresponding mean profile for L-mode.
Outside of $\rho=0.2$, the normalised temperature gradient 
decreases more slowly than the L-mode gradient.

The most important control variable in predicting the H-mode temperature
shape is the heating fraction. As in L-mode, ICRH heating creates much more
peaked profiles than does the equivalent heating with NBI. Comparing
the response functions for $f_H(\rho)$ for H-Mode (Fig.~\ref{HTmu}),
and for L-mode (Fig.~\ref{LTmu}) shows that the two curves have similar
shapes, but that the H-mode function is less peaked in the H-mode case.
This indicates that the L-mode profile shape is more sensitive to the type 
of heating than is the ELM-free H-mode

Table \ref{HTPAR} displays the model selection criteria 
as we increase the number of control variables in the model.  
When only the heating fraction is used, the PAR statistic is
.0650. If the effective charge is added, the PAR value is
.0632, while using the line average density instead of $Z_{eff}$
yields a worse value of .0640. We believe that this difference
is large enough that we choose $Z_{eff}$ as the second control variable.
We should caution that the errors in $Z_{eff}$ are appreciably larger than 
in the other control variables. Also, $Z_{eff}$ is a spatially varying
quantity and we are parameterizing the impurity distribution using one
chordal measurement. A final caution is that $f_Z(\rho)$ is largest
at the plasma edge, corresponding to the edge temperature increasing 
with increasing $Z_{eff}$. Since a higher edge temperature will produce
impurities, our results could be interpreted as saying that broader
temperature profiles create more impurities. 
Our fit only shows that higher $Z_{eff}$ correlates with broader profiles,
and we cannot comment on the causality issue. 
Even if the edge temperature is creating the $Z_{eff}$, our fit shows how
the entire temperature profile shape response to the resulting impurity
influx.

The last column of Table \ref{HTPAR} shows that the fit does not improve
when a third parameter is added. 
Thus the  sequential selection procedure suggests 
the two control variable  model:
\begin{equation}\label{E5}
T(\rho)/ \Tbr = \mu_0(\rho) Z_{eff}^{f_{Z} (\psb)} \exp\left(f_H(\rho)H_{fr}
\right) \ .
\end{equation}
Since $f_{Z} (\psb) \approx - f_H(\rho)$, our profile shape depends
almost exclusively on $H_{rf} - \ln[Z_{eff}]$.
Figure~\ref{HTdat} plots our predictions versus the data for the 
NBI heating-only profile  with the 
smallest $Z_{eff}$ concentration (1.07) 
and for the ICRH-only profile with the largest $Z_{eff}$ concentration (5.43).
The control variable, $H_{rf} - \ln[Z_{eff}]$ is -.068 for the NBI discharge
and -.692 for the ICRH discharge; thus our model predicts that 
this particular dirty ICRH  discharge should be broader than the clean
NBI discharge. Our prediction is in clear agreement with the data,
indicating that the $Z_{eff}$ effect can negate the difference in
heat type.

The line average temperature has been regressed using a  sequential selection
procedure. Our best fit yields
\BEQ \label{HTbr}
\Tbr = .649 I_p^{1.22\pm.17} P_L^{.29\pm.08} (\nbr)^{-.23\pm.13} 
\exp((.24\pm .08)H_{fr}) \ .
\NEQ
The corrected RMSE is 19.8 \%, again indicating that the 
line average temperature may not follow a power law scaling.
Similarly to L-mode, the line average temperature depends on $\nbr$ and $P_L$
only through the ratio of $P_L/\nbr$ (to good approximation), corresponding
to the power per particle.
The line average temperature has a weak density  dependence $(\nbr)^{-.246}$.
If we assume that the total plasma energy scales proportionally to
$\nbr_e \Tbr_e$, then Eq.~(\ref{HTbr}) implies that the plasma energy
would scale as $\nbr^{.75}$. This result is consistent with a gyro-Bohm
scaling, but differs from the commonly observed confinement
dependence of $\tau_E \sim \nbr^{.15}$. 
Part of this discrepancy may be explained
because fast particle confinement degrades with increasing density.

The power scaling in Eq.~(\ref{HTbr}) is consistent with a confinement scaling
of $\tau_E \sim P_L^{-.7}$, in agreement with most scaling expressions.
The line average temperature  has an even stronger current dependence
than the corresponding confinement time has (typically$^{\cite{ITER89}}$
$\tau_E \sim I_p^{.85}$ to $I_p^{1.0}$.)
We note that $q_{95}$ and $I_p$ vary  roughly half as much in the
H-mode database as in the L-mode database. As such, the uncertainty in
the scalings with $q_{95}$ and $I_p$ is  larger in H-mode.
Nevertheless, the $I_p$ variation is sufficiently large to show that
the H-mode density profile shape
has a highly significant $I_p$ dependence. 

\section{DISCUSSION}

Our parameterizations of the JET normalised temperature profiles
fit the LIDAR  measurements with a mean predictive  error of
.075 for Ohmic, .063 for H-mode and .087 for L-mode.
The corresponding density fits have PAR values of
.056 for Ohmic, .056 for H-mode and .061 for L-mode.
Thus our fits accurately describe the profiles in our
database, and they may be used in the modeling of these
discharges as a proxy for the real data.

The model selection criteria determine which control variables are
most important. Table \ref{SumTab} 
summarises which variables modify the
profile significantly in each regime. 
In both L-mode and H-mode, the heating type dominates the profile shape
due to either power deposition effects or sawtooth stabilization.
The Ohmic density profile depends only on  the Murakami parameter,
and thus the profile broadening appears to be related to radiation effects.
The  L-mode density profiles
depend on the line average density  with broader profiles corresponding
to higher densities, while the H-mode density depends primarily on the
plasma current. 
If a control variable is not used in our parameterization, 
it means that the omitted variable was not necessary to include
in the modeling.

Physics trends are often discernible from the profile fit.
One clear line of research is to explain the observed parametric dependencies
as summarised in Table \ref{SumTab}. 
The most tantalizing physics results are the shape dependencies on
heating type and the $\nbr/B_t$ dependence of the Ohmic density.


We caution that our results are based on a  subset of
the JET discharges. 
Our existing data
contains a variety of different discharges which are representative
of typical JET operating regimes. We fully expect that our findings
are representative of the usually observed JET profile dependencies.
Our H-mode data set is identical to the
data set which we later used in the initial  DIII-D-JET H-mode comparison
in collaboration with Dave Schissel$^{\cite{SSIR}}$.
In this article,
we include  slightly different sets of control variables than
in the combined fit. The parameterization given here is
optimised to give an accurate fit to the JET data set alone.
In Ref.~\cite{SSIR}, the control variables and resulting parameterization
were optimised to fit both machines simultaneously. 


We stress that our log-additive profile fits are only an approximation of
reality. Our philosophy is ``All models are wrong; some are useful.''
We believe that our power law expressions fit the data well and can be
used as summary of existing JET results and as a benchmark for new results.
These expressions summarise the observed profile shape as a function of the
control variables.

Our profile parameterisations may be used directly in analysis codes.
By fitting many discharges simultaneously, we reduce the 
discharge-to-discharge variation at the cost of making systematic
model errors. These model errors are typically small in the
parameter region where the data is taken, but can be large
when the parameterization is extrapolated into new regimes.
Since our fits can be evaluated in real time, they can be used in
plasma control systems. In  the near future, they will  be evaluated
on-line as part of  the JET LIDAR diagnostic. Thus, one can quickly
compare a particular discharge  temperature and density shapes
with ``standard'' JET results.

\

\cl{ACKNOWLEDGMENTS}

\noindent
G.~Cordey's support and encouragement are gratefully acknowledged.
We thank  
C.~Gowers for his many helpful suggestions.
The final revision of this paper has benefited significantly
from our new collaboration with  Dave Schissel.
We thank the referees for their useful comments.

KI's work was supported by the U.S. Department of Energy Grants
No.DE-FG02-92ER54157 and 86ER-53223.
KSR's work was supported by the U.S. Department of Energy Grants
DE-FG02-86ER-53223 and 91ER54131.

\ \\

{\bf APPENDIX: PREDICTIVE ERROR ESTIMATION}

In the present article, we use the predictive absolute residual (PAR)
criterion to select which terms to 
measure the goodness of fit and determine which control variables 
influence the profile shape: 
\begin{equation}\label{E00}
{\bf PAR} = 
\sum_{i,j} {|T_i ( \rho_j )- \hat{T} (\rho_{j},{\bf u}_i )| \over
{N - 2 \times \rm \# \ of \ free\  parameters }} \ ,
\end{equation}
where $T_i ( \rho_j)$ is the measurement of the
normalized temperature of the $i$th profile
at the $j$th measurement location, and
 $\hat{T}$ is the corresponding fitted value of the 
normalised temperature (or normalised
density) as a function of the vector of control variables, ${\bf u}$.
In the denominator of Eq.~(\ref{E00}), $N$ is the total number of measurements.
Since we are using  smoothing splines, the number of free parameters
decreases as the smoothness penalty increases.
In the appendix of Ref.~\cite{IRS}, 
our definition of the number of free parameters
in a spline fit is given.
The factor of two 
accounts for the increased difficulty in predicting
new data instead of fitting the existing data.
As the number of free parameters increases, the denominator tends
to make the PAR value increase. This effect tends to counterbalance the
improvement in the residual error from adding more free parameters.

The PAR statistic is simply the least absolute value analog of
the Rice  criterion, $C_R$, which we used in our previous 
work$^{\cite{RI,IRS}}$:
\begin{equation}\label{A13}
 C_R \equiv {1 \over {N - 2 \times \rm \# \ of \ free\  parameters }}
\sum_{i,j} {|\ln[T]_i ( \rho_j )- \widehat{\ln[T]} (\rho_{j},{\bf u}_i )|^2 \over
\sigma_{i,j}^2} \ 
\ , 
\end{equation}
where $\sigma_{i,j}^2$ is the variance of the measurement $\ln[T]$
for the $i$th profile and $j$th measurement location.
In the appendix of Ref.~\cite{IRS}, 
we give a derivation the Rice criterion. 

The PAR criterion is our estimate of the expected absolute error
in predicting the normalised profiles of new data taken under similar
operating conditions. In theory, $(N-2p) PAR *\sigma^2$ has approximately
a $\chi^2$ distribution with  $(N-2p)$ degrees of freedom
where $p$ is the effective number of degrees of freedom in the model. 
Given the large
amount of data, $N-2p \sim 2000$, and the significant departures of the
data from the simple assumptions of the statistical model, we do
not trust the PAR statistic to determine when to truncate the model.
Instead, we stop adding terms to the model when several different
control variables yield a similar reduction in the PAR criterion.

The PAR statistics sums the {\em absolute values} of the fit errors
while the Rice criterion
sums the $squares$ of the fit errors on the logarithmic scale
normalized to the standard deviation of the measurement error. 
Standardizing the residual errors to the measurement errors is optimal
when the fit error is proportional to the measurement error.
In practice, our fit errors are only weakly correlated with
the measurement error size. 
A reasonable hypothesis is that the systematic error
portion of the fit error is uniform and uncorrelated with the measurement
error size.  

By using the absolute value, the PAR statistic is more robust
in the sense that it is less sensitive to a small number
of points which fit poorly.
As is standard in robust statistics, we use the degree of freedom
correction for Gaussian statistics, but replace the sum of squares
estimate of the residual variance with a robustified analog. 
The PAR
statistic corresponds to the visual quality of fit while the Rice
criterion tends to measure the fit error near the plasma edge where
the fit error is the largest on the relative scale.
When one or two profiles have large residual fit errors, the
Rice criterion will depend sensitively on the residual errors of
these profiles due to its quadratic weighting. In contrast, the PAR
statistic  does not emphasise the most poorly fitting
profiles due to its linear weighting, and therefore  is more robust.
The Rice criterion can be evaluated much more rapidly,
and we use it for optimizing the smoothing parameters. We use PAR to
determine which covariants to include in the model.


In our earlier analyses$^{\cite{RI,IRS}}$, we minimise Eq.~(\ref{A13}) 
with respect to the smoothing parameters. 
This yields a smoothing parameter, $\lambda_R$, which is
nearly optimal with regard to predictive mean square error.
However, the resulting estimated curves often have spurious wiggles which
we do not believe are actually present. To remove these wiggles,
we increase the smoothing parameter by a factor of $\ln(N)$.
The logarithmic factor suppresses artificial wiggles with asymptotic
probability one$^{\cite{RiedelPCC}}$.

An older (obsolete) statistic is $\chi^2$, which replaces the denominator of
($ N - 2 \times$ \#  of  free  parameters) with
($N -  \#$  of  free  parameters), and thereby
corresponds to the mean square error per degree of freedom. 
The $\chi^2$ statistic is useful in optimizing the fit
to existing data, while the Rice criterion 
minimises the predictive error for new data.
The factor of two in the denominator of $PAR$ and $C_R$ 
accounts for the greater difficulty in predicting new data than in fitting
existing data.
This factor of two in the denominator of $C_R$ results in smoother models 
and fewer variables in the model.


\ \\

\ \\

\ \\

\np



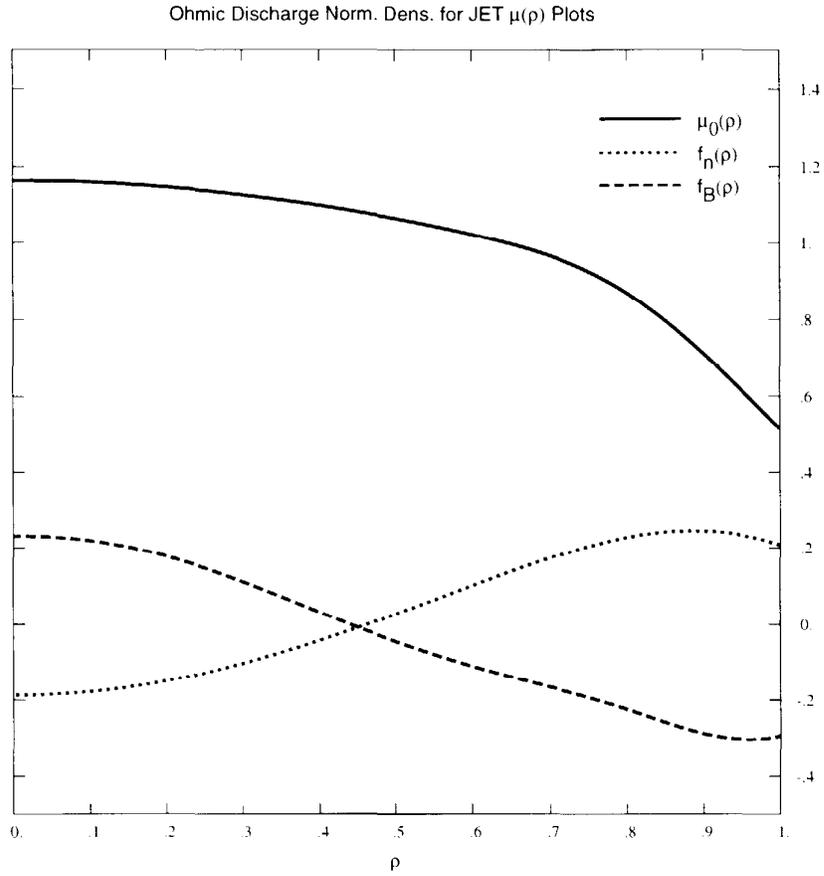

Figure 1

FIG.1. Fit functions for the normalised Ohmic density profiles. Here $\mu_0(\rho)$ gives the fitted profile at $\bar{n}/B_t =$ 0.765. The profile becomes hollow with increasing $\bar{n}/B_t$.

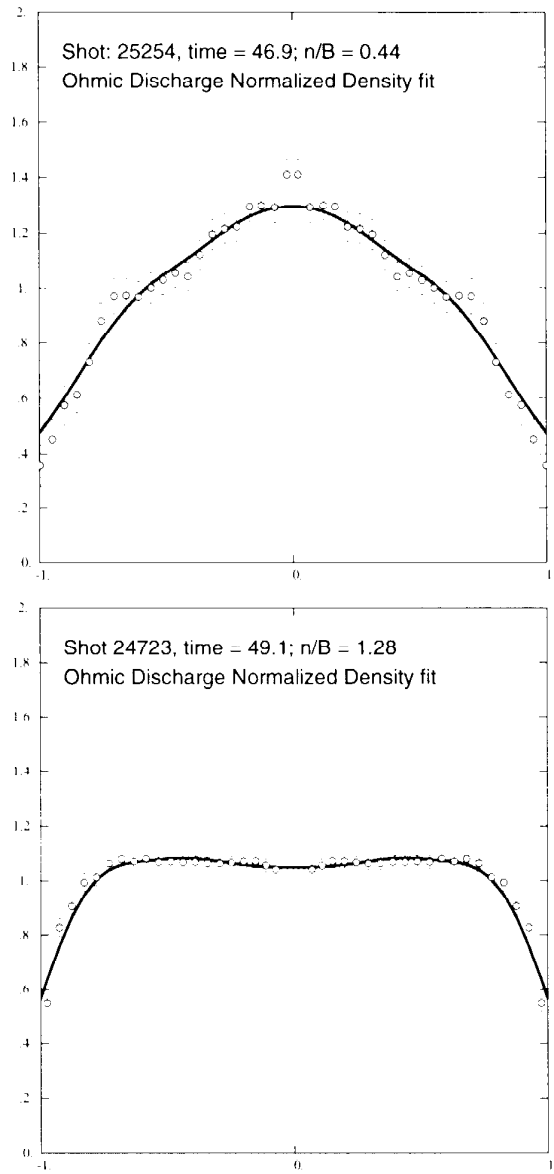

*FIG.2. Normalised data versus the predicted Ohmic density profile. The broader curve (JET discharge #24723 at time = 49.1 sec) has the largest value of $\bar{n}/B_t$ (1.28) in the data set. The narrower curve (JET discharge #25254 at time = 46.9 sec) has the smallest value of $\bar{n}/B_t$.*

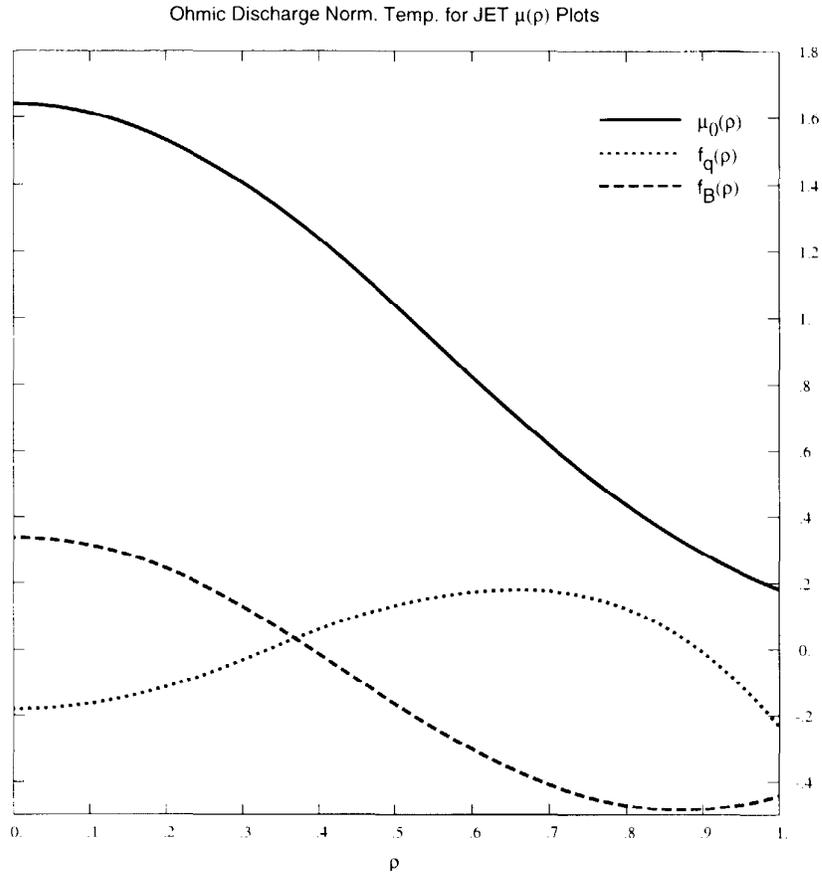

Figure 3

*FIG.3. Fit functions for the normalised Ohmic temperature profiles. Here $\mu(\rho)$ gives the fitted profile at $q_{95} = 4.968$, $B_t = 2.72\ T$. The profile becomes peaked with increasing $q_{95}$ or decreasing $I_p$.*

23

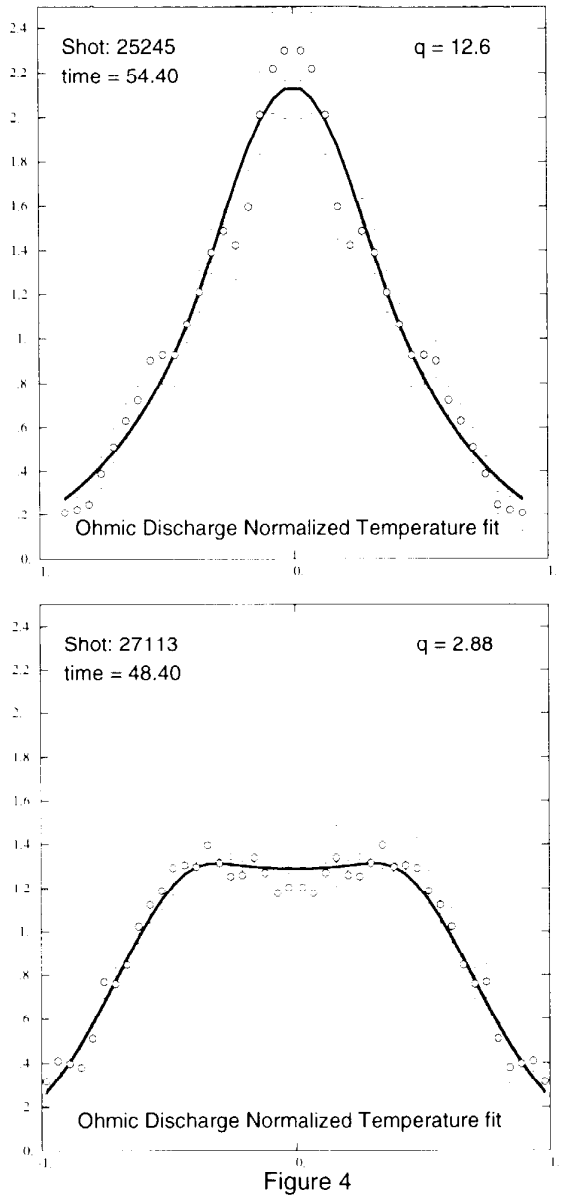

*FIG.4. Normalised data versus the predicted Ohmic temperature profile. The broader curve (JET discharge #27113 at time = 48.4 sec) has the largest value of $q_{95}$ (2.88) in the data set. The narrower curve (JET discharge #25245 at time = 54.4 sec) has the smallest value of $q_{95}$ (12.6) in the data set.*

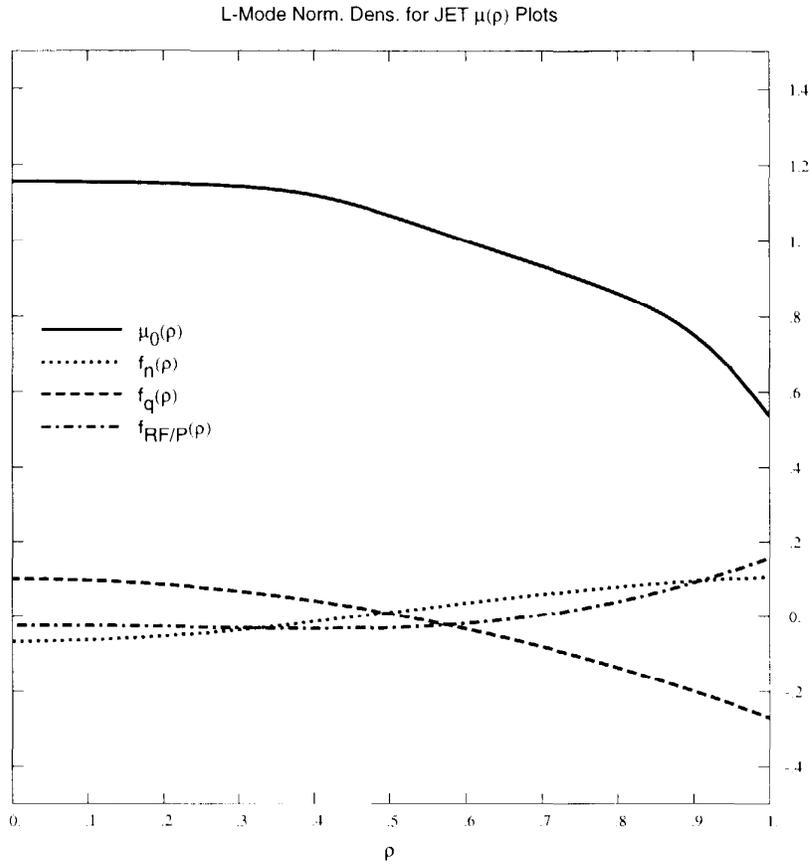

Figure 5

*FIG.5. Fit functions for the normalised L-mode density profiles. The canonical profile, $\mu(\rho)$, is the predicted profile at $\bar{n} = 2.09\ 10^{19}\ m^{-3}$, $q_{95} = 5.066$, and $H_{fr} = .471$. Increasing $\bar{n}$ results in hollow profiles.*

25

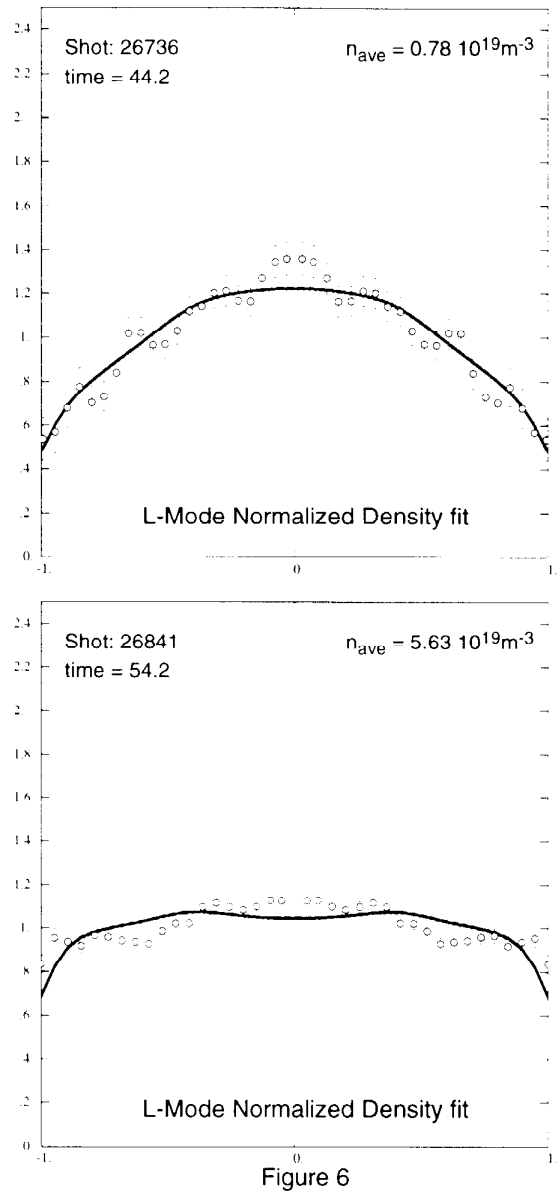

*FIG.6. Normalised data versus the predicted L-mode density profile. The broader curve (JET discharge #26841 at time = 54.2 sec) has the largest value of $\bar{n} = 5.63\ 10^{19}\ m^{-3}$) in the data set. The narrower curve (JET discharge #26736 at time = 44.2 sec) has the smallest value of $\bar{n} = 0.78\ 10^{19})\ m^{-3}$ in the data set.*

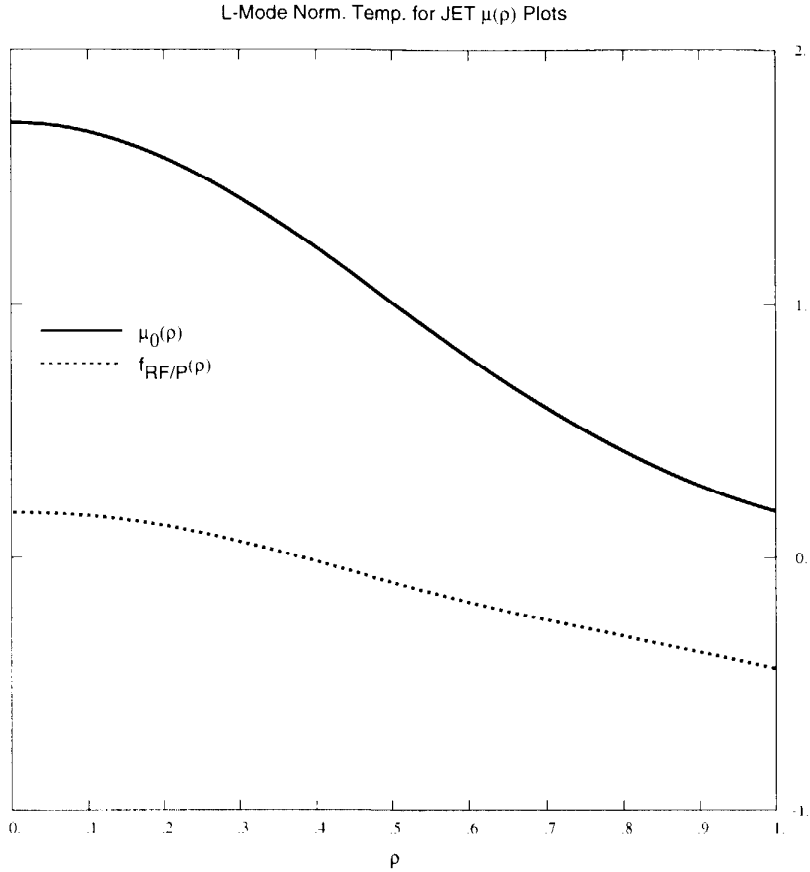

Figure 7

FIG. 7. Fit functions for the L-mode temperature. The canonical profile, $\mu(\rho)$, has a Gaussian shape with the temperature gradient decreasing outside of $\rho = 0.8$. Only the heating type fraction is important in determining the profile shape. As the percentage of ICRH power increases, the profiles become more peaked, as seen from the inferred $H f_H(\rho)$ curve. Here $\mu(\rho)$ corresponds to the predicted profile when the heating fraction is .471, corresponding to the database mean.

27

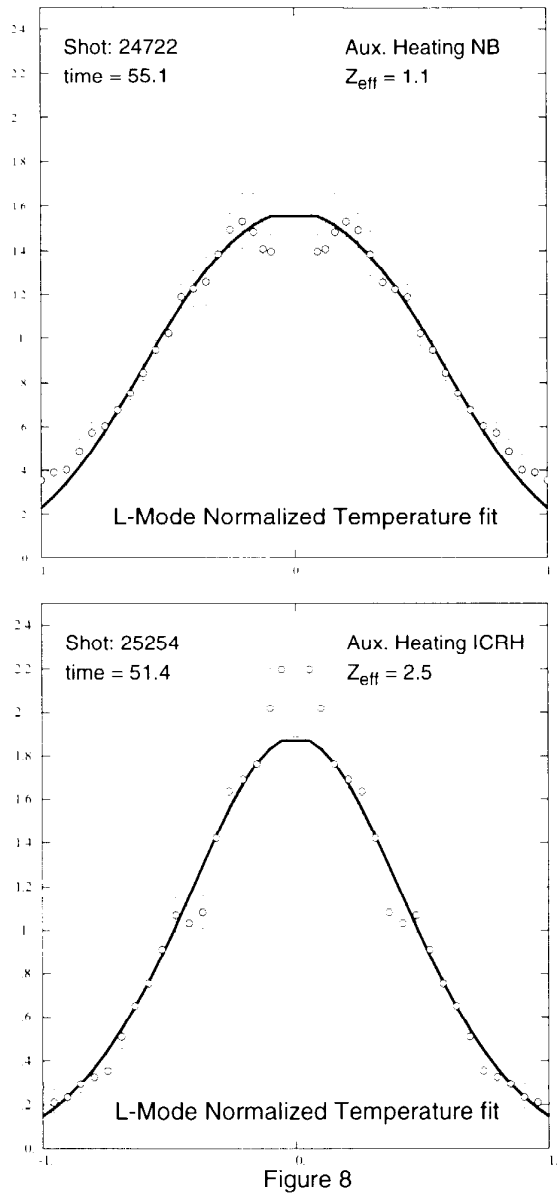

FIG. 8. *Normalised data versus the predicted L-mode temperature profile. The broader curve (JET discharge #24722 at time = 55.1 sec) has the smallest value of $Z_{eff}$ (1.1) for a NBI discharge (in the data set. The narrower curve (JET discharge #25254 at time = 51.4 sec) has the largest $Z_{eff}$ (2.5) of the ICRH discharges in the data set.*

28

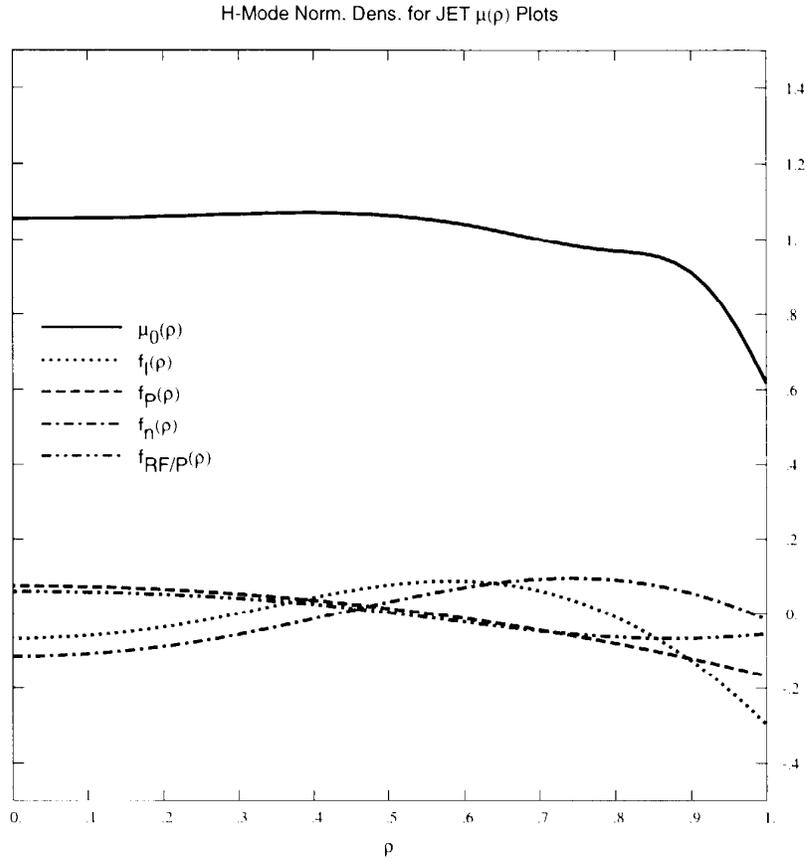

Figure 9

FIG. 9. Fit functions for the normalised H-mode density profiles. The canonical density profile is nearly flat. Here $\mu_0(\rho)$ is our predicted profile at $I_p = 2.66$ MA, $\bar{n} = 3.70\ 10^{19}\ m^{-3}$, $P_L = 6.32$ MW and the heating fraction, $H_{fr} = .368$. Increasing the plasma current broadens the profile and steepens the gradient at the edge.

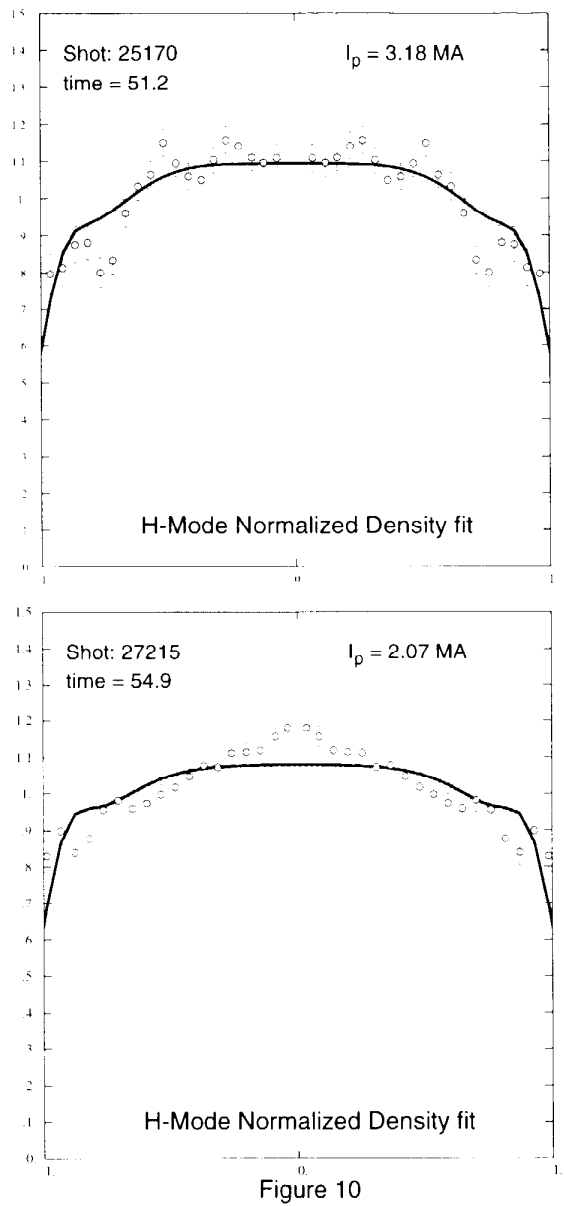

FIG. 10. Normalised data versus the predicted H-mode density profile. The broader curve (JET discharge #25170 at time = 51.2 sec) has the largest value of $I_p$ (3.18 MA) in the data set. The narrower curve (JET discharge #27215 at time = 54.9 sec) has the smallest value of $I_p$ (2.07) in the data set. Large values of $I_p$ result in hollow profiles with sharp drops near the plasma edge.

30

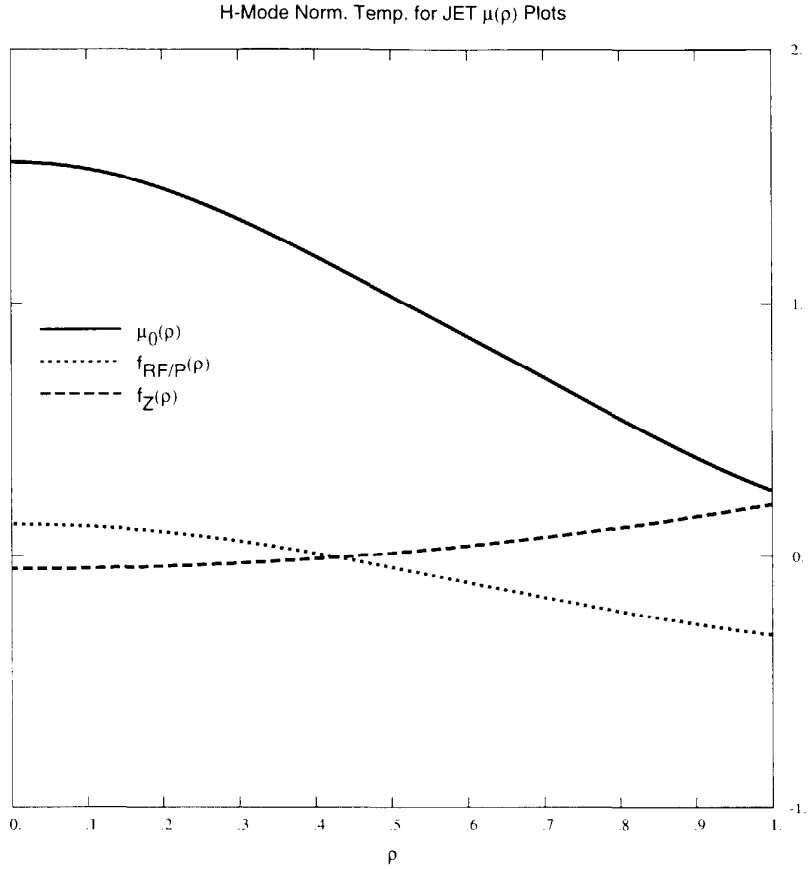

Figure 11

FIG. 11. Fit functions for the H-mode temperature. The canonical profile $\mu_0(\rho)$, is broader than the L-mode profile. Here $\mu_0(\rho)$ is our predicted profile at a heating fraction, $H_{fr} = .368$ and $Z_{eff} = 2.328$. Although the heating fraction is the most important control variable, $f_H(\rho)$ is less peaked than the corresponding function for L-mode. Since $f_- f_Z(\rho) \approx -f_H(\rho)$, the profile shape depends only on $H_{fr} \cdot \ln[Z_{eff}]$.

31

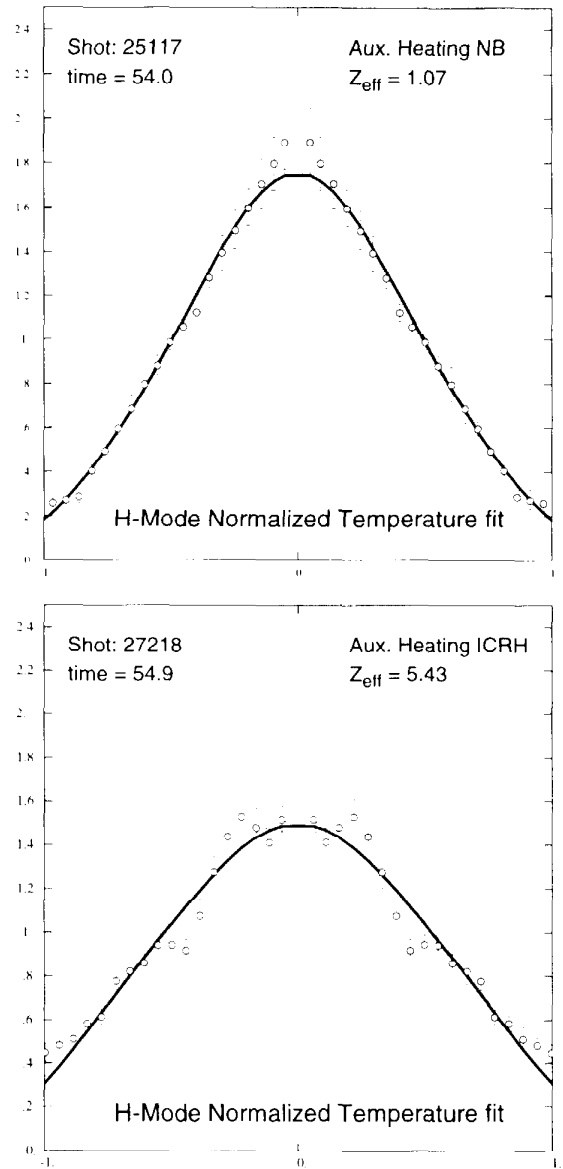

FIG. 12. Normalised data versus the predicted H-mode temperature profile. The narrower curve (JET discharge #25117 at time = 54.0 sec) has the largest value of $Z_{eff}$(1.07) for a NBI discharge in the data set. The broader curve (JET discharge #27218 at time = 54.9 sec) has the largest $Z_{eff}$ (5.43) of the ICRH discharges in the data set. The key variable, $H_{fr}$ - $ln[Z_{eff}]$, is .692 for the ICRH discharge and -.068 for the NBI discharge.



\newpage
{\cl{\bf TABLES}} 
\vspace{.25in}

\begin{table}
\begin{center} 
\begin{tabular}{|r|c|c|c|c|}
\hline
Var & mean &min & max &std dev\\ 
\hline \hline 
${\overline n}$    & 2.20  &  1.27    & 3.69   & 0.74\\
$q_{95}$     & 5.48  &  2.88    & 12.6   & 2.86\\
$I_p$        & 2.80  &  0.97    & 5.25   & 1.13\\
$B_t$        & 2.76  &  1.30    & 3.22   & 0.46\\
$\kappa$     & 1.44  &  1.30    & 1.75   & 0.122\\
$a$          & 1.16  &  1.05    & 1.19   & 0.040\\   
$R$          & 2.92  &  2.83    & 3.01   & 0.047\\
$Z_{eff}$  & 2.10  &  1.20    & 3.35   & 0.60\\
Volt         & -.35  & -1.12    & .914   & 0.66\\
\hline
\end{tabular}
\end{center}
\

\caption{
Ohmic database summary: Average, minimum,  maximum and 
standard deviation of each of the control variables.
\label{ODB} }
\end{table}

\ \\

\begin{table}
\begin{center}
\begin{tabular}{|r|c|c|c|c|}   
\hline 
         Var & mean &min & max &std dev \\
\hline
$\bar{n}$    & 2.31  &  0.78    & 5.63   & 1.04     \\
$q_{95}$     & 5.35  &  3.38    & 16.58   & 2.34     \\
$I_p$        & 3.07  &  1.04    & 4.87   & 0.83   \\
$B_t$        & 2.62  &  1.44    & 2.89   & 0.44     \\
$P_{L}$    & 4.81  & 0.46     & 14.55 & 2.73                             \\
$\kappa$     & 1.67  &  1.50    & 1.83   & 0.10    \\
$Z_{eff}$  & -  &  1.07    & 7.61   & -     \\
\hline 
\end{tabular}
\end{center}

\vspace{.12in}

\caption{L-mode database summary:
Average, minimum,  maximum and
standard deviation of each of the engineering control variables.
\label{LDB}}
\end{table}

\ \\

\begin{table}
\begin{center}
\begin{tabular}{|r|c|c|c|c|}
\hline 
         Var & mean &min & max &std dev \\
\hline
$\bar{n}$    & 3.85  &  1.878    & 6.73   & 1.19     \\
$q_{95}$     & 5.19  &  3.10    & 6.99   & 1.21     \\
$I_p$        & 2.70  &  2.07    & 3.18   & 0.45   \\
$B_t$        & 2.53  &  1.43    & 2.90   & 0.41     \\
$P_{L}$    & 8.00  & 0.81     & 12.81  & 2.90 \\
$\kappa$     & 1.73  &  1.59    & 1.84   & 0.061    \\
$a$          & 1.04  &  0.98    & 1.13   & 0.038     \\
$R$          & 2.89  &  2.77    & 2.97   & 0.043     \\
$V_{loop}$         & -.16  & -1.12    & 0.17   & 0.22     \\
 $Z_{eff}$  & 2.63  &  1.07    & 7.61   & 1.27     \\
\hline 
\end{tabular}
\end{center}

\caption{H-mode database summary:
Average, minimum,  maximum and
standard deviation of each of the engineering variables.
\label{HDB} }
\end{table}

\

\begin{table}
\begin{center}
{\bf Ohmic Density Goodness of Fit Table}

\

\begin{tabular}{|l|c|c|c|c|} \hline 
& 1 spline & 2 spline & 3 spline & 1 spline   \\ \cline{2-5}
 & PAR & PAR &  PAR & PAR\\ \hline
$\bar{n}$  & \underline{.0612} & spline    & spline & .0565  \\
$q_{95}$ & .0633 & .0593  & .0566 & .0561\\
$I_p$ &.0668 & .0613  & .0564 &.0560 \\
$B_t$ & .0681 & \underline{.0567}  & spline & .0567   \\
$\kappa$ & .0687 & .0611  & .0568 & .0562 \\ 
$Z_{eff}$ & .0674 & .0604  & .0561 & .0556\\
$\Tbr$ & .0683 & .0597  & .0563 &.0555\\
$\bar{n}/B$ & \underline{.0561}  & --- & --- & \underline{spline} \\
\hline 
\end{tabular}
\end{center}

\vspace{.2in}

\caption{Goodness of fit of different log-additive models
for the normalised Ohmic density.
The ``Murakami'' parameter, $\nbr/B_t$, reduces the PAR statistic
the most. In the third column, we compare two variable
models using $\nbr$ in each case. 
The best two variable model in column 3  is $\nbr$ and $B_t$. 
The fourth column shows that
adding a third control variable doesn't improve the fit
significantly.
In the final column, we compare two variable
models using $\nbr/B_t$ in each case. Using only the  ratio
of $\nbr/B_t$ outperforms using both $\nbr$ and  $ B_t$ separately
because fewer degrees of freedom are used in the fit.
\label{ODPAR}}

\end{table}

\ \\

\begin{table}
\begin{center}
{\bf Normalised Ohmic Temperature Goodness of Fit Table}

\

\begin{tabular}{|l|c|c|c|} \hline 
& 1 spline & 2 spline & 3 spline   \\ \cline{2-4}
 & PAR & PAR &  PAR \\ \hline
$\bar{n}$  & .0969 & .0818  & .0776 \\
$q_{95}$ &\underline{.0812} &  spline   & spline   \\
$I_p$ &  .0863 & {.0779}  & .0748 \\
$V_{loop}$ &  .0995 & .0818  & .0763 \\
$B_t$ & .0933 & \underline{.0755}  & spline   \\
$\kappa$ & .0963 & .0790  & .0737 \\
$Z_{eff}$ &  .0888 & .0789  & .0760 \\
$\Tbr$ & .0961 & .0804  & .0761 \\
\hline 
\end{tabular}
\end{center}

\caption{Goodness of fit of different log-additive models
for the normalised Ohmic temperature.
The last row shows that the profile shape does not depend
on the line average temperature, $\Tbr$, thereby justifying
our normalisation.
The edge safety factor, $q_{95}$, reduces the PAR statistic
the most. In the second column, we compare two variable
models using $q_{95}$ in each case. The best two variable
model is $q_{95}$ and $B_t$. The final column shows that
adding a third control variable doesn't improve the fit
significantly. Since $\kappa$ varies little, the $\kappa$ 
dependence is probably due to hidden variables associated with
machine operation. 
\label{OTPAR}}

\end{table}

\vspace{.2in}

\ \\

\begin{table}

\begin{center}
{\bf L-mode Density Goodness of Fit Table}  

\

\begin{tabular}{|l|c|c|c|c|} \hline 
Contr & 1 spline & 2 spline & 3 spline & 4 spline  \\ \hline{2-5}
Var. & PAR  & PAR & PAR  &  PAR \\ \hline \hline 
$q_{95}$ & .0667 & \underline{.0619} & spline & spline \\
$\bar{n}$ & \underline{.0633} & spline & spline & spline \\
$I_p$ & .0714 & .0634 & .0623 & .0610 \\
$B_t$ & .0705 & .0637 & .0621 & .0609 \\
$\kappa$ & .0668 & .0627 & .0607 & .0605 \\
$\bar{n}/B_t$ & .0641 & .0635 & .0619 & .0609 \\
$P_{L}$ & .0714 & .0626 & .0613 & .0603 \\
$H_{fr}$ & .0688 & .0628 & \underline{.0606} & spline \\
$Z_{eff}$ &  .0672 & .0639 &0623 & -- \\
$\Tbr $ & .0706 & .0632 & .0610 & .0606 \\
\hline 
\end{tabular}
\end{center}

\vspace{.2in}
 
\caption{Goodness of fit of different log-additive models
for the normalised L-mode density.
``Spline'' variables are included in each run in that  column.
We then add the
variable that reduces  the criterion the most.
The line average density improves the fit the most, followed
by adding the edge $q$ and then the heating fraction.
Adding a fourth control variable only slightly improves
the goodness of fit.
\label{LDPAR}}
\end{table}


\ \\

\begin{table}
\begin{center}
{\bf Temperature L-mode Goodness of Fit Table}

\

\begin{tabular}{|lc|c|c|c|c|} \hline 
Contr & 1 spline & 2 spline & 3 spline & 4 spline 
  \\ \hline \hline 
Var & PAR &  PAR & PAR & PAR \\ \hline
$q_{95}$ &  .1034 & .0873 & .0871 & .0872  \\
$\bar{n}$ & .1025 & \underline{.0868} & spline & spline \\
$I_p$ & .1020 & .0875 & .0874 & .0874 \\
$B_t$ & .1012 & .0874 & .0872 & .0873 \\
$\bar{n}/B_t$ & .1035 & .0869 & .0872 & .0875  \\
$\kappa$ & .0960 & .0874 & .0866 & .0865  \\
$P_{L}$ & .1033 & .0873 & \underline{.0868}  & spline \\
$H_{fr} $ & \underline{.0869} & spline & spline & spline \\
$Z_{eff}$ & .102   & .0868 &  .0871 & -- \\
$ \Tbr     $ & .102 & .0873 & .0870 & .0873 \\
\hline 
\end{tabular}

\end{center}

\

\vspace{.2in}
\caption{Goodness of fit of different log-additive models
for the L-mode temperature.
The L-mode temperature depends almost exclusively on the type of heating.
Adding more traditional control variables does not improve the 
fit. Even using four control variables gives a fit 
roughly comparable to the fit using only the heating fraction.
The last row shows that the profile shape does not depend
on the line average temperature, $\Tbr$. Thus the size and shape
of the temperature have little correlation.
\label{LTPAR}}
\end{table}

\begin{table}
\begin{center}
{\bf H-mode Density Goodness of Fit Table}

\

\begin{tabular}{|l|c|c|c|c|} \hline 
Contr & 1 spline & 2 splines & 3 splines & 4 splines \\ \cline{2-5}
Var  & PAR &  PAR & PAR &PAR \\ \hline \hline
$\bar{n}$  & .0645  & .0617 &\underline{.0584} & \underline{.0563}\\
$q_{95}$ & .0625 & .0614  & .0596 &.0585  \\
$I_p$ & \underline{.0619} & spline & spline & spline  \\
$B_t$  & .0650 & .0613 & .0597 & .0587 \\
$Z_{eff}$ & .0647 & .0612 & .0599 & .0584 \\
$P_{L}$ & .0633 & \underline{.0597} & spline & spline \\
$H_{fr}$ & .0638 & .0603 &\underline{.0583} & spline \\ 
\hline 
\end{tabular}

\end{center}
\vspace{.15in}

\caption{
Goodness of fit of different log-additive models
for the normalised H-mode density. The plasma current is the
most important control variable followed by the loss power.
At the third stage, both $\nbr$ and the heating fraction
are equally effective in reducing the PAR statistic.
  This table illustrates
a difficulty of sequential variable selection:
Occasionally, there is no clear cutoff in the number or choice
of terms to include. We select the four-variable model because
the PAR value continues to decrease and the $(I_p,P_L,\nbr,H_{fr})$
model is clearly superior to other four variable models.
\label{HDPAR}}
\end{table}

\begin{table}
\begin{center}
{\bf H-mode Temperature Goodness of Fit Table}

\ \\

\begin{tabular}{|l|c|c|c|c|} \hline \hline
Contr & 1 spline & 2 spline & 3 spline  & 4 spline  \\ \cline{2-5}
Var. & PAR  & PAR & PAR  &  PAR   \\
\hline \hline
$  \bar{n}$ & .0738 & .0640 & .0630 & spline \\
$q_{95}$ & .0735 & .0655 & .0639 & .0634  \\
$I_p$ & .0739 & .0645 & .0641 & .0636  \\
$B_t$ & .0700 & .0646 & .0638 & .0632  \\
$\kappa$ & .0697 & .0652 & .0636 & .0633  \\
$Z_{eff}$ & .0728 & \underline{$.0632$}  & spline & spline \\
$P_{L}$ & .0738 & .0648 & .0636 & .0625  \\
$H_{fr}$ & \underline{$.0650$} & spline & spline  & spline \\
$\Tbr$ &    .0726 & .0562 & .0640 & .0634 \\
\hline 
\end{tabular}
\end{center}

\caption{
Goodness of fit of different log-additive models
for the ELM-free H-mode temperature.  The most important control variable
is the the heating fraction of ICRH power.
Adding $Z_{eff}$ to the model results in a significant improvement in
fit. No significant reduction in the PAR statistic occurs when a third
control variable is added.
\label{HTPAR}
}\end{table}


\ \\

\newpage


\begin{table}

\begin{center}

{\bf Scaling Expressions for Profiles}

\

\begin{tabular}{llll} \hline 
 & Ohmic & L-Mode & H-Mode \\ \hline
density & $\ntil(\rho )    
=  \mu(\rho)  ({\bar{n} \over B_t} )^{f_{n/B} (\rho )}$
&
$\ntil(\rho ) 
=  \mu(\rho)(\bar{n})^{f_n (\rho )}\ q_{95}^{f_{q} (\rho )}$
&
$\ntil(\rho ) 
= \mu(\rho) I_p^{f_I (\rho )}P_L^{f_{P} (\rho )} (\nbr)^{f_{n} (\rho )} $ \\
shape &$ $
 & $\ \ \  \times \exp[f_H(\rho )H_{fr}]$
& $\ \ \  \ \ \times
\exp[f_H(\rho )H_{fr}] $ \\ 
temperature & $\Ttil(\rho ) 
=  \mu(\rho)q_{95}^{f_q (\rho )}B_t^{f_B (\rho ) }  $
&
$\Ttil(\rho) =  \mu(\rho)\times$
&
$\Ttil(\rho )=  \mu(\rho)\times$
\\
shape 
&   &
$\exp[f_H(\rho )H_{fr}] $ &
$\exp[f_H(\rho )H_{fr}] Z_{eff}^{f_Z(\rho)}  $ \\ 
Line average
& $ \Tbr \sim I_p^{.64}B_t^{.54} (\bar{n} )^{-.31} $
&
$\Tbr \sim B_t^{.97} P_{L}^{.36}( \bar{n})^{-.38}$ 
&
$ \Tbr\sim I_p^{1.22}  P_{L}^{.29}(\nbr)^{-.253}$
\\
temperature
&   &  &
\ \ \ \ \ $\exp[.24 H_{fr}]  $ \\ \hline 
\end{tabular}
\end{center}
\caption{Summary of scaling dependencies for profiles.
The normalised profiles are denoted by 
$\ntil(\rho ) \equiv n(\rho ) / \nbr$ and 
$\Ttil(\rho ) \equiv T(\rho ) / \Tbr$.
\label{SumTab}
}\end{table}

\end{document}